# Examining persistence of European open repository infrastructure and its diffusion in the scholarly record

George Macgregor
Information Services
University of Glasgow
https://orcid.org/0000-0002-8482-3973

Joy Davidson
Digital Curation Centre
University of Glasgow
https://orcid.org/0000-0003-3484-7675

**Abstract**

This article seeks to determine the extent to which the principle of persistence is observed by repositories and the organizations that operate them. We also evaluate the impact that negative repository persistence levels may be having on the scholarly record. We do this by interrogating and combining data about European repositories from several repository registries and web scraped sources, including the Internet Archive's Wayback Machine, thereby creating a unique dataset of historic repository locations and their OAI-PMH endpoints. We then use this data as the basis for text mining CORE, a vast corpus of scholarly outputs, to determine the extent to which impersistent European repository content has permeated the scholarly literature. Our findings indicate that over a fifth of European repositories (> 20%) could be classified as 'dead', with an even greater proportion (> 40%) of the machine interfaces associated with these repositories similarly dead. Problematically, our analysis indicates that circa 12,000 unique scholarly works cite, refer to, or actively used this repository content – amounting to circa 19,000 unique repository locations – all of which are now dead and unretrievable from their stated resource location. Partly owing to limitations in available repository registry data and the existence of 'zombie' repositories, there are reasons to conclude that the total number of scholarly works referring to dead repository content is far higher. We also find evidence of dead repository content entering the current scholarly record, a phenomenon we describe as 'dead on arrival' referencing. We consider the implications of these observations, proffer explanations, and propose possible policy interventions to address the issue of repository persistence. Our dataset also enables us to make several observations about the nature of impersistent repositories, their profile, and their decay rate.

Correspondence should be addressed to Author Name, Postal Address. Email: email@ddress

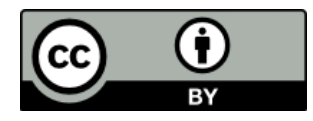

# Introduction

The expansion of open scholarly repositories within research institutions has significantly reshaped the way research is conducted and shared (DANS. et al., 2022; Suber, 2012). The increasing accessibility of the content exposed by repositories enables a wider audience to engage with digital research resources, fostering an open research environment with its many advantages. A need for 'trustworthy' repositories has emerged as an essential enabler of a functioning and healthy scholarly ecosystem, in which research resources demonstrate FAIRness (Collins et al., 2018). This is because repositories have become key components of global open scholarly infrastructure, supporting the sharing of publications, data, and other resources.

However, a growing crisis in the persistence of scholarly research resources has emerged. Despite the long-standing recognition that stable addressing on the web is critical to resource persistence (Berners-Lee, 1998), many scholarly resources are disappearing (Jones et al., 2016; Klein et al., 2014). This crisis coalesces with wider concerns about the reproducibility of scholarly research, the ability of scholars to verify findings, and broader questions about trust and integrity within scholarship more generally (Baker, 2016; Brembs, 2018; Gertler & Bullock, 2017; Helgesson & Bülow, 2023). Technical mechanisms devised to better support the persistence of scholarly resources have also been found to be unreliable, exacerbating these concerns (Klein & Balakireva, 2022).

Scholarly repositories, in their various permutations, have long been considered reliable nodes within global open scholarly infrastructure. From the beginning they were acknowledged as 'essential infrastructure for scholarship' by better enabling discovery of digital content, but also by delivering persistent access to the scholarly record and supporting the long-term stewardship of digital resources (Lynch, 2003). Repositories have consequently tended to be operated by research organizations with expertise in digital object management, notably university research libraries. Although such organizations purport to observe digital object management best practice, this does not always appear to be reflected in practice. Emerging evidence suggests that some repository instances, and the organizations that operate them, are demonstrating poor repository management by moving content in unpredictable ways, mismanaging repository URIs, or retiring repositories entirely (Bamgbose et al., 2025; Macgregor, 2025; Strecker et al., 2023). The explanations for why this is happening can be complex (Bamgbose et al., 2025; Rothfritz et al., 2025). But suffice to state that rather than mitigating the persistence crisis, many such organizations appear to be contributing to it, thereby compromising the integrity of the scholarly record and hindering content discovery, research citation, verification, and reproducibility — all of which are essential to academic trust.

This article seeks to determine the extent to which the principle of persistence is observed by repositories and the organizations that operate them. Better understanding this problem space is what motivates our study. We evaluate the impact that negative repository persistence levels may be having on the scholarly record. We do this by interrogating and combining data about European repositories from several repository registries and web scraped sources, and then by using this data as the basis for text mining CORE (Knoth et al., 2023), a vast corpus of scholarly outputs, to determine the extent to which impersistent European repository content has permeated the scholarly literature. We consider the implications of these observations, proffer explanations, and propose possible policy interventions to address the issue of repository persistence. To assist us in our observations and analysis we define the notion of 'dead' and 'zombie' repositories. Our dataset also enables us to make several observations about the nature of impersistent repositories, their profile, and their decay rate. Though we cannot always determine the reason why a repository may be dead, our research contributes to an improved understanding of the threat to the digital scholarly record which exists through scholarly infrastructure that is otherwise perceived to be persistent. It also complements the wider, growing body of evidence examining 'link rot', 'reference rot', and the persistence crisis more generally.

# Background and Context

## URI Persistence

The growing fragility of some web resources and the need for their URIs to be managed responsibly was drawn to wider attention by Tim Berners-Lee (1998). Berners-Lee noted that: “When someone follows a link and it breaks, they generally lose confidence in the owner of the server [and] are frustrated from accomplishing their goal”. Berners-Lee articulated a series of principles to which content publishers should adhere to ensure persistence in resources. These included the allocation of URIs that ‘should not change’ and were stable over long time (i.e. 20-200 years), and that those creating URIs are mindful of their ‘design’ such that they support long-term management (Berners-Lee, 1998). The centrality of persistent URIs to W3C initiatives such as the Semantic Web and Linked Data meant that Berners-Lee’s principles were later formalised through W3C recommendations, including guidelines on ‘Cool URIs for the Semantic Web’ (Sauermann & Cyganiak, 2008). Similarly, as nodes within open scholarly infrastructure and as scholarly data hubs, an adherence to responsible URI management within repositories has long been considered essential (Hockx-Yu, 2006; Nelson & Allen, 2002; Shreeves & Cragin, 2008). The importance of Cool URIs within the publication of FAIR data on repositories and the associated knowledge graphs arising from such publication has therefore also recently emerged (Thalhammer, 2024), highlighting the continued relevance of creating stable, secure, and persistent URIs.

## Persistence and the Scholarly Web

Despite an early understanding of the problems impersistent URIs present for digital libraries (Phelps & Wilensky, 2000; Shreeves & Cragin, 2008) and the integrity of scholarship (Carnevale & Aronsky, 2007; Ducut et al., 2008; Eysenbach & Trudel, 2005; Russell & Kane, 2008; Spinellis, 2003), research continues to uncover concerning levels of ‘reference rot’ within the scholarly record. In a deep analysis of scholarly literature within the Science, Technology, and Medicine (STM) domains, Klein et al. reported that one fifth of links to URI references suffered reference rot (Klein et al., 2014). In subsequent work, members of the same research group further revealed significant levels of ‘content drift’ occurring within URI references (Jones et al., 2016), with > 75% of references pointing to web content that had altered since its original citation. This line of research was further refined by Massicotte and Botter (2017) who studied the level of reference rot and content drift within doctoral theses and dissertations served by a large university repository. Their findings were proportionally consistent with Klein et al. (2014), with 23% of thesis references demonstrating link rot, though their detection of content drift was slightly lower, at circa 50%. Results from these studies, and others like them (Martin-Segura et al., 2022), contribute to wider concern about the integrity of the scholarly record and the ability of future scholars to verify assertions or findings reported within the scholarly literature.

Theoretical work performed by Romero (2025) has concluded that trust is both a critical and desired effect of improved transparency in scholarship. A lack of transparency arising from, for example, the inability to verify scientific assertions in the literature is therefore highly damaging to trust in research findings and scholarship more generally.

Attempts to prevent link or reference rot occurring in the first place is a complex socio-technological challenge, requiring longer-term behavioural change across scholarship and digital publishing more generally (Silva, 2021). This includes improved digital literacy among scholars, some of whom struggle to cite digital scholarly objects accurately (Van de Sompel et al., 2016), or correctly identify commonplace persistent identifiers, such as digital object identifiers (DOIs) (Macgregor et al., 2023). For this reason, research has sought to mitigate the impact of reference rot by proposing innovative interventions. In an early study testing the retrieval of ‘decayed’ online citations from works published in the early 2000s, Dimitrova and Bugeja (2007) reported that only circa 53% of the citations they tested were retrievable using the Wayback Machine, the most reliable of the systems they evaluated. They concluded that the Wayback Machine was too unreliable to depend upon for scholarship and noted concerns about how decaying citations 'threatened' the scientific method.

However, as the Wayback Machine has evolved, so too have the possibilities for harnessing machine methods to access its data to resolve citation gaps, where they might occur. A rich stream of research work to enable a framework for time-based content negotiation has therefore delivered the Memento Protocol (Jones, Klein, Sompel, et al., 2021; Klein, Balakireva, et al., 2019; Klein, Shankar, et al., 2019; Zhou et al., 2015). Initially a technical proposal (Van de Sompel et al., 2009), the Memento Protocol (Van de Sompel et al., 2013) has since been embedded within some scholarly tools to solve the content negotiation associated with the reference rot phenomenon (Mahanama et al., 2022). 'Time travel' and archiving within Memento-aware web browsers has also been demonstrated (Mabe et al., 2021, 2022). Subsequent work has sought to improve the strength of hyperlinks by proposing 'Robust Links', an approach that seeks to archive web resources at the point of their citation and include machine-actionable annotations to the link such that the resource can be accessed in both a live or archived state (Jones, Klein, & Van de Sompel, 2021; Klein et al., 2018).

## Persistence: Repositories as Scholarly Infrastructure

Work to ameliorate reference rot, such as through the Memento Protocol and Robust Links, has arisen largely in response to the ephemeral nature of some web resources (Klein et al., 2018; Van de Sompel et al., 2009). Open scholarly repositories – whether they be institutional or subject-based, or publication focused or data centric – are usually contrasted with the wider ephemeral web as reliable and persistent components of scholarly infrastructure. From their inception these repositories were considered as essential infrastructure, not only because they enhanced the discovery of digital scholarly content and supported the goals of the open access movement (Suber, 2012); but also because they ensured persistent access to scholarly materials and supported the long-term stewardship of digital resources (Crow, 2002; Francke et al., 2017; Hockx-Yu, 2006; Lynch, 2002, 2003). For these reasons many open scholarly repositories have been operated under the auspices of research libraries. This has typically been explained by their operational overlap with other types of digital library or archive (Burns et al., 2013). A recent survey of European repositories conducted by COAR[1], LIBER[2], OpenAIRE[3], and SPARC Europe[4] confirms the continued centrality of the university library to repository management (Shearer et al., 2023).

Repositories are beneficial services in which digital content can be deposited, stored, and managed, thereby affording considerable discovery advantages to authors and organizations (Dong & Tay, 2023; Kelly, 2023; Thompson & Hoover, 2023). Generally, they also remain conducive to URI persistence insofar as they are typically built using open-source software solutions (e.g. DSpace, Invenio, EPrints, etc.) and have usually been designed to support responsible URI management concepts (Macgregor, 2025). This will typically entail a URI syntax that predictably describes the scheme → subdomain → domain → folder/context or accession identifier (Thalhammer, 2024), resulting in simple, predictable, and less brittle URIs.

It might be expected that repository content which uniquely underpins research, and ergo the scientific method, might therefore demonstrate greater persistence. However, a study performed by Strecker et al., using data from the re3data data repository registry, found that repository retirement was not uncommon, reporting that 6.2% of research data repositories indexed in the registry had been retired (Strecker et al., 2023). Strecker and colleagues describe the negative implications for the scholarly record while also noting the lack of communication about 'repository shutdown events'. In an exploratory pilot study of UK-based repository persistence, Macgregor (2025) indicated that 45% and 31% of UK repository OAI-PMH endpoints and repository home URL locations respectively were 'dead', with a combination of repository response errors reported (e.g. 404 or 400 responses, NXDOMAIN, etc.). Perhaps for these reasons scholars are increasingly encouraged to make use of trustworthy digital repositories (TDRs) to share their work. Organizational mismanagement of repositories, however, significantly undermines trustworthiness.

[1] Confederation of Open Access Repositories (COAR): https://coar-repositories.org/

[2] Ligue des Bibliothèques Européennes de Recherche – Association of European Research Libraries (LIBER): https://libereurope.eu/

[3] Open Access Infrastructure for Research in Europe (OpenAIRE): https://www.openaire.eu/

[4] Scholarly Publishing and Academic Resources Coalition Europe (SPARC Europe): https://sparceurope.org/

On the theme of 'trust' as it relates to data repositories, Yakel et al. (2024) found that scholars' trust in reusing a data repository was predicted by the observable behavioural actions of the managing organization (Yakel et al., 2024). In other words, integrity — central to the concept of trust — arises not by simply stating that there is, for example, a commitment to ensuring data preservation over time, but by behaving in a manner that demonstrates adherence to such a commitment. The results described by Strecker et al. (2023) therefore represent an unhelpful development in the wider, positive need to ensure scholarly resources are deposited, disseminated, shared, and responsibly maintained over time. It also undermines important pan-organizational initiatives, such as IMPACT-REPO, which seeks to safeguard Europe's repository infrastructure by "ensuring research remains accessible, trusted, and reusable" (Shearer et al., 2025).

As evidence emerges of their potential 'untrustworthiness', the reliability of repositories as nodes within open scholarly infrastructure has recently been scrutinized. A systematic review performed by Bamgbose et al. (2025) sought to understand the extent to which there was 'trustworthiness' of the repositories operated by research libraries. As a systematic review, Bamgbose et al. could not arrive at a measure of untrustworthiness but they were able to conclude that, while most repositories were deemed trustworthy, there were widespread concerns in the literature about the inadequate technical support or resource often afforded to repositories, as well as the financial constraints and librarian 'disinterest'. An overreliance by organizations on short-term funding arrangements for repository operations (sometimes between 2-5 years), as well as a shortage of suitably qualified human resource, has been observed to contribute to the challenges of longer-term repository management, threatening trustworthiness (Jouneau et al., 2025). Such an observation is arguably compounded by the importance of transparency in supporting trustworthiness (Romero, 2025) and the related importance of visible and understandable repository policies, which may be absent (Jouneau et al., 2025) — since their transparency at least enables scholars to make informed decisions about where to deposit scholarly work (Yakel et al., 2013).

Rothfritz et al. (2025) conducted a wider systematic review designed to surface greater understanding of the challenges confronting institutional repositories. The financial sustainability of repositories was again identified as a particular challenge in all global regions, with low acknowledgement of the unintended consequences of staffing constraints and resource limitations of repository mismanagement. In support of the IMPACT-REPO initiative, Shearer et al. (2025) tacitly acknowledge the challenging environment confronting repository infrastructure. They reiterate the need for organizations, and the wider scholarly community, to recognize their strategic importance and invest to secure their future, including superior observance of FAIR principles, greater use of persistent identifiers, and the upskilling of staff to ensure repository management 'excellence' (Shearer et al., 2025).

## Research Motivation: Dead and Zombie Repositories

Our study is motivated by better understanding the size and nature of an emerging repository persistence crisis. Rather than functioning as reliable nodes for the dissemination and management of scholarly research objects and thereby helping to address the persistence and reference rot pervading the scholarly web at large, there is emerging evidence that some repositories are exacerbating the problem. We assess how low levels of repository persistence may be affecting the integrity of the scholarly record. To achieve this, we aggregate and analyse data on European repositories from multiple sources, including various repository registries and web-scraped datasets. Europe is home of the largest continental concentration of repositories (Nazim & Bhardwaj, 2023). It therefore provides a sufficiently large indicator of global repository persistence while also reducing the computational overhead associated with text mining all global literature. Our approach enables us to construct a unique dataset that captures historical repository locations and their OAI-PMH endpoints. Using this dataset, we then conduct text mining on CORE (Knoth et al., 2023)—a comprehensive collection of scholarly outputs—to evaluate the degree to which unstable European repository content has been referenced or incorporated into the scholarly literature.

To assist in our discussion and analysis we define two types of impersistent repository: 'dead' repositories, and 'zombie' repositories.

A 'dead' repository can be defined as follows:

- A once active repository which had a history of successfully serving content, but which suddenly becomes unavailable over the Internet, either through active retirement or neglect. Such repositories will therefore tend to demonstrate domain name registration errors (i.e. NXDOMAIN), report HTTP 404 responses, etc. These dead repositories will not display evidence of Cool URI management, normally because the repository service has been terminated or has been poorly maintained.

A sub-category of dead repository is a 'zombie' repository. A zombie repository can be defined as:

- A once active repository which had a history of successfully serving content before suddenly becoming unavailable. But, unlike a dead repository, is often 'reanimated' under a new domain, typically using different software and serving content from alternative locations. Limited Cool URI management is usually displayed (e.g. reusing existing URIs, formally redirecting users / machines to new location, etc.). Users / machines are also unaware that the location of the repository has changed, and persistence is therefore disrupted.

It is important to note that zombie repositories are also dead ones, hence they are deemed a sub-category; but we highlight this distinction here because, as we shall see, such zombie approaches to repository management are equally as disruptive to content persistence as a terminally dead repository might.

# Methods: Data Collection and Preparation

## Repository Registry Data

To test the persistence of repositories, it is first necessary to consult an inventory of their existence. Such repository registries exist though their data can come with limitations (Walk, 2023). Baglioni et al. (2025) have proposed a data model to improve interoperability of scholarly repository registries, given their diversity and disparate data models. To build a suitable registry dataset for our study, we curate a dataset from three sources using a combination of API interrogation and web scraping:

1. Open Directory of Open Access Repositories (OpenDOAR): OpenDOAR[5] data are exposed as JSON objects via the Jisc Open Policy Finder API (formerly the Sherpa API)[6]. Using the API, we queried and extracted registry data pertaining to European repositories, capturing key registry data, including repository name, home URL, OAI-PMH endpoint location, and country code.

[5] Open Directory of Open Access Repositories (OpenDOAR): https://opendoar.ac.uk/
[6] Jisc Open Policy Finder: https://openpolicyfinder.jisc.ac.uk/

2. Registry of Open Access Repositories (ROAR): Though ROAR[7] does not provide a public API it is instead possible to perform a full JSON export of its registry data on European repositories directly from its UI. Data captured overlaps with OpenDOAR but includes key registry data, such as repository name, home URL, OAI endpoint, etc.

Both OpenDOAR and ROAR are principally concerned with curating a contemporaneous registry of repositories and do not model change arising from zombie repositories. Walk (2023) has, for example, reported staleness and gaps in OpenDOAR's registry data. We therefore combine our OpenDOAR and ROAR data with the IAR.

3. Institutional Archive Registry (IAR): The 'Institutional Archives Registry' was the precursor service to ROAR and was retired in the mid-2000s. We use a snapshot from the Wayback Machine, captured by the Internet Archive on 13 June 2006, at which point 750 repositories were registered on the IAR. Using web scraping techniques, we extracted 301 repositories identified as falling within our European scope and scraped associated text and hyperlink data about each of the registry entries. Key registry data elements were gathered using this process, though country codes had to be assigned programmatically.

All three registry sources categorize repository types differently. OpenDOAR records five repository categories only. These include institutional, disciplinary, aggregating, governmental, and undetermined. ROAR supports twelve, though many of these types could be interpreted as subcategories of those provided by OpenDOAR. To enable aggregate observations of the repository type data, we inspected indicative examples of the ROAR and IAR repository types and imposed mappings to the OpenDOAR types, as in Table 1.

Data from all three sources were merged, cleaned, and deduplicated, with redundant data removed. Superfluous URL query parameters, erroneously present in some repository registry entries, were removed to identify the next, best stable non-parametered URL. The resulting dataset – shared as part of this research paper – represents a unique inventory of active and historic repositories, including dead and zombie repositories. It forms a benchmark from which it is possible to make assessments about repository persistence over time and the scholarly impact of impersistence, where it exists.

**Table 1.** Mapping of repository types from ROAR and IAR registry data to the OpenDOAR repository type vocabulary.

| IAR repository type | ROAR repository type | OpenDOAR repository type |
|---|---|---|
| Research institutional or departmental | Research institutional or departmental | Institutional |
| Not applicable | Research multi-institution repository | Institutional |
| Research cross-institution | Research cross-institutional | Institutional |
| Not applicable | Subject | Disciplinary |
| e-journal/publication | e-journal/publication | Undetermined |
| e-theses | e-theses | Institutional |
| Database | Database | Aggregating |
| Not applicable | Research data | Institutional |
| Not applicable | Open data | Aggregating |

[7] Registry of Open Access Repositories (ROAR): https://roar.eprints.org/

| Not applicable | Learning and teaching objects | Institutional |
|---|---|---|
| Demonstration | Demonstration | Undetermined |
| Not applicable | Web observatory | Undetermined |
| Other | Other | Disciplinary |

## HTTP(S) Response Data

To identify dead repositories within our dataset we deployed a script to gather HTTP status request codes for every repository domain URL and its associated OAI-PMH endpoint. OAI-PMH remains an 'essential characteristic' of repository functionality (Confederation of Open Access Repositories, 2022), currently providing the only standard machine protocol for repository content discovery and data exchange (Knoth et al., 2023) OAI-PMH facilitates interoperability across heterogeneous repositories, enables metadata harvesting and aggregation, and remains a 'cornerstone' protocol of repository workflows (Rubin et al., 2026). Its importance is reflected in repository registry records such as OpenDOAR. The absence of a functioning repository OAI-PMH endpoint, combined with a repository domain URL, is therefore a significant signifier of repository health.

HTTP responses were logged against the repository entries within our registry dataset. Common HTTP response codes are widely documented by the IETF and are available for reference (Fielding et al., 2022). To optimize response accuracy, we employed GET HTTP requests as the authoritative retrieval method and as a more reliable technique for obtaining response metadata and behaviour (Fielding et al., 2022). These requests were stateless and used distributed cloud infrastructure (Google Cloud) to best model repository service behaviour under disparate network conditions (e.g. production-scale routing behaviour). This is significant since many services dynamically tailor responses based on IP-derived geolocation and content delivery network (CDN) routing. The approach used therefore provides a broader view of operational behaviour than a single fixed client location would allow.

It should be noted that many registry data in our dataset described repositories prior to the widespread adoption of HTTPS. HTTP to HTTPS redirects (HTTP 301 response) to essentially the same location – such as http://example-repository.uni.de to https://example-repository.uni.de – were therefore parsed by our script as a 200 response.

Though data pertaining to repository uptime was not something explicitly collected during our experiments, it is possible to report that many repositories demonstrated erratic availability during the period of data collection. This could possibly be associated with recent 'bad bot' behaviour by Large Language Model (LLM) agents (Knoth, 2025; Sherrick & Pino Navarro, 2024). However, other repositories lacked valid SSL certificates, limiting browser or agent access. This meant that manual checks and re-tests of repositories were often necessary to verify whether unavailability was merely temporary, intermittent, or permanent.

## Repository 'Date of Decease' Data

We queried the Wayback Availability JSON API[8] to capture an approximate 'date of decease' for repositories found to be returning unsatisfactory HTTP response codes. This API is partially based on the Memento Protocol and serves archived snapshot data on the last available website archive, including `archived_snapshot_URL` and associated timestamp. Data retrieved from the API were logged against the repository entries within our registry dataset.

[8] Wayback Availability JSON API: https://archive.org/help/wayback_api.php

## CORE: Mining the Scholarly Literature

The final part of our data collection concerned mining the scholarly literature to understand the extent to which any impersistent European repository content has entered the scholarly record. We use CORE (Knoth et al., 2023) as the basis for this text mining. CORE represents a vast open corpus of scholarly outputs, providing access to the world's largest collection of open publications. These publications are harvested and aggregated from the global network of repositories and journal titles, enabling user discovery through search but also scientific discoveries through text and data mining (TDM) techniques. CORE has become an important feature of the open scholarly infrastructure landscape (Jefferies et al., 2022) and is increasingly a platform upon which value-added services are developed (Knoth et al., 2023).

CORE provides a sophisticated API to expose the scholarly resources it has harvested and enriched, using its own CORE API Query Language[9]. We deployed a script to query version 3.0 of the CORE API to determine the extent to which impersistent European repository content has permeated the scholarly literature. This script sought to mine CORE's corpus for scholarly works that cite, refer to, or actively used dead repository content.

JSON responses from the CORE 'Works' API were processed to log key bibliographic data elements. This included work `title`, `authors`, `documentType`, `doi`, `identifiers`, `id` (an internal CORE identifier), `oaiIds` (OAI identifiers associated with repositories), `yearPublished`, and `depositedDate`. Data arising from the `fullText` field were parsed and any in-text references to dead repository content were extracted, using the URL prefix of offending repositories, with a wildcard. Extracted URLs underwent normalization and validation. It should be noted that not all works in CORE have full text available for TDM but many include data enrichments performed by CORE (e.g. reference data from CrossRef, etc). This is a limitation we shall return to later in the paper.

All data were captured and the dataset compiled during the month of June 2025.

# Results

## Dead Repositories

Data gathered and merged from our repository registry sources identified 3,751 repositories falling within our European scope. The characteristics of these repositories are summarized in Table 2, the vast majority of which were 'institutional' (~82%) with 'disciplinary' accounting for the next largest repository type (~12%). Overview results arising from our HTTP response data to identify dead repositories can also be summarized within Table 2. A total of 824 repositories were found to be deceased, equivalent to 22% of all repositories within our dataset, leaving 2,927 active repositories. Here we note that institutional repositories were the most common type of dead repository ($n$ = 620), accounting for > 75% of all repositories found to be deceased. As a highly represented repository type within our dataset, this is perhaps unsurprising; however, it should be noted that as a proportion of all repositories within our dataset, this indicates that one fifth (> 20%) of all institutional repositories are now deceased. The strong presence of disciplinary repositories ($n$ = 147) should be noted as being almost double the combined total of other dead repository types ('aggregating', 'governmental', 'undetermined').

**Table 2.** Summary data by repository type within the analysed dataset and summary observations of dead repositories. Based on domain name HTTP response status data.

| Repository type | Total by type | Total by type (%) | Dead count | Dead as % of total dead count | Dead as % of repository type | Dead as % of all repository types |
|---|---|---|---|---|---|---|

[9] CORE API (3.0.0): https://api.core.ac.uk/docs/v3

| institutional | 3081 | 82.1 | 620 | 75.2 | 20.1 | 16.5 |
|---|---|---|---|---|---|---|
| disciplinary | 445 | 11.9 | 147 | 17.8 | 33.0 | 3.9 |
| aggregating | 104 | 2.8 | 25 | 3.0 | 24.0 | 0.7 |
| governmental | 59 | 1.6 | 10 | 1.2 | 16.9 | 0.3 |
| undetermined | 62 | 1.7 | 22 | 2.7 | 35.5 | 0.6 |
| **Totals** | 3751 | 100 | 824 | 100 | – | 22 |

The HTTP response data overview in Table 2 includes assessments of repository death based on the home domain name of a repository. Our method sought similar data based on the OAI-PMH endpoints associated with repositories in our dataset since support for this protocol is central to our understanding of repositories, their discovery, and data re-use (Macgregor, 2023). For reasons described later in the results and discussion, the response status of an OAI-PMH endpoint can also be an indicator of a zombie repository. Data arising from analysis of OAI-PMH responses is set out in Table 3.

A total of 1643 OAI-PMH endpoints within our dataset were found to be dead, with > 78% (*n* = 1289) of these dead endpoints belonging to institutional repositories. This is equivalent to > 34% of all repositories within the entire dataset. Combined with the dead endpoints identified for all repository types, well over one third of repository OAI-PMH endpoints (circa 44%) within our dataset can be reported as dead. We can observe significant proportional disparities between repository types. Institutional and disciplinary repositories contribute to the overall volume measure of dead repositories, but within 'governmental' and 'undetermined' types we can observe circa 58% and 52% dead endpoints respectively. For example, our dataset identified 59 governmental repositories, of which 34 appeared to have a dead OAI-PMH endpoint. These repository types may be small relative to our entire dataset but generate massive proportional results within their respective category type.

**Table 3.** Summary data by repository type within the analysed dataset and summary observations of dead repositories. Based on HTTP response status of associated OAI-PMH endpoints.

| **Repository type** | **Total by type** | **Total by type (%)** | **Dead count** | **Dead as % of total dead count** | **Dead as % of repository type** | **Dead as % of all repository types** |
|---|---|---|---|---|---|---|
| institutional | 3081 | 82.1 | 1289 | 78.5 | 41.8 | 34.4 |
| disciplinary | 445 | 11.9 | 243 | 14.8 | 54.6 | 6.5 |
| aggregating | 104 | 2.8 | 45 | 2.7 | 43.3 | 1.2 |
| governmental | 59 | 1.6 | 34 | 2.1 | 57.6 | 0.9 |
| undetermined | 62 | 1.7 | 32 | 1.9 | 51.6 | 0.9 |
| **Totals** | 3751 | 100 | 1643 | 100.0 | – | 43.8 |

National profiles emerging from repositories in our dataset can be summarized, as in Table 4. Here data are organized by two-letter country codes (as per ISO 3166-1 alpha-2) and ranked by dead repository count (domain). 41 countries are represented. Unsurprisingly, larger countries can be observed to have the larger total repository counts but, perhaps disappointingly, also the larger count of dead repositories. The United Kingdom (gb) can be reported as the leader based on count, with circa 24% of its 528 repositories and circa 41% of associated OAI-PMH endpoints found to be dead. Percentage measures of dead repositories once again suggest significant impact for small categories (e.g. Luxembourg (lu), Belgium (be), and Bulgaria (bg)), where the proportional impact of dead repositories was found to be as high as 80%. This is clearer to note in Figure1, in which at times the tail of dead repositories by country (at domain level) suggests an inverse proportional increase within some regions; however, regression analysis revealed no statistically significant association ($R^2$ = .006, $p$ = .629).

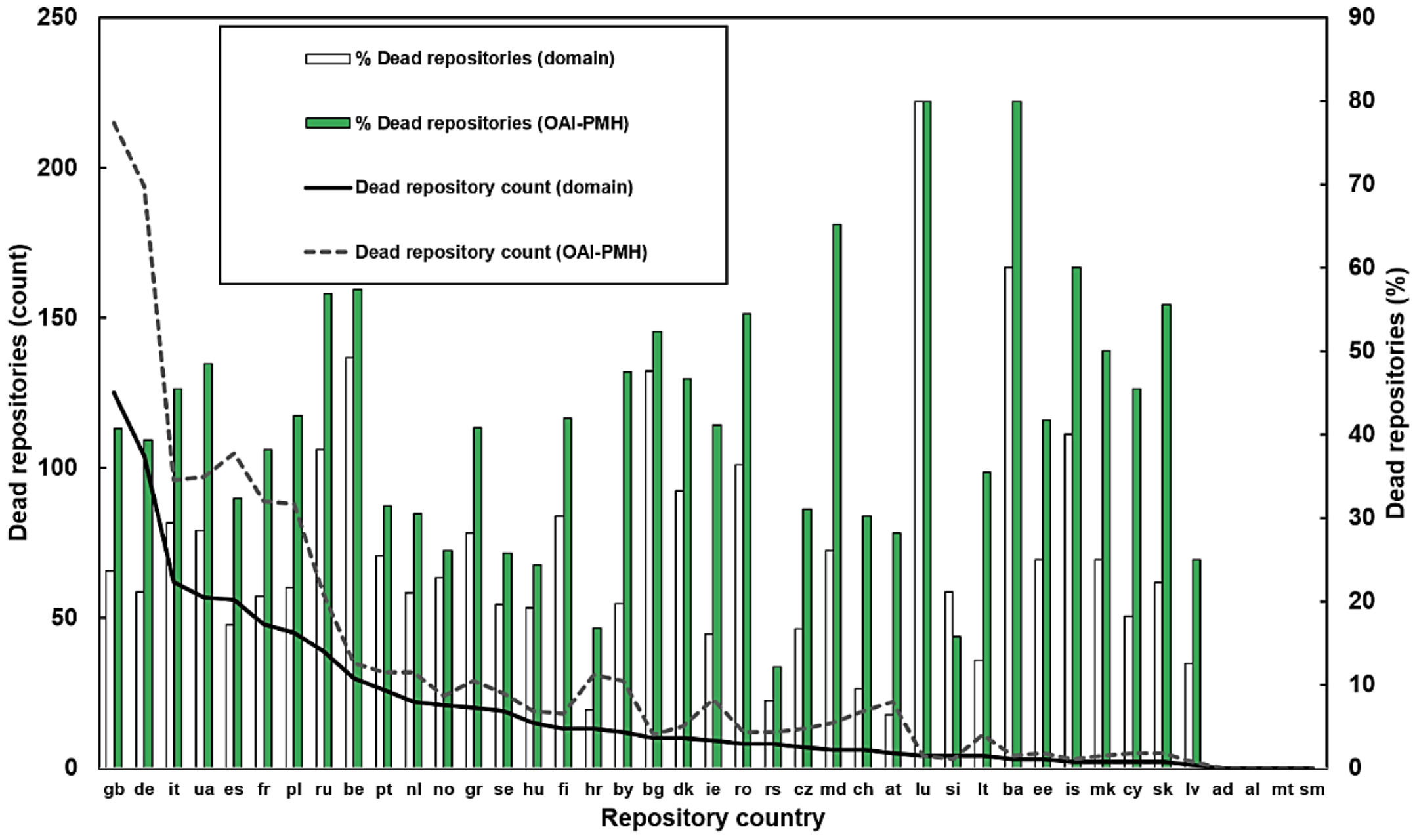

**Figure 1.** Number of dead repositories at domain and OAI-PMH level by country and as a percentage of total repositories within European countries.

The general disparity between dead repository domains and their associated OAI-PMH endpoints is observable from Tables 2 and 3. But we can observe the differential between dead repository domains and endpoints by country in Table 4. Both the United Kingdom (gb) and Germany (de) report the highest number of dead repository domains and OAI-PMH endpoints, but also the largest differential between the two. The significance of this differential is to be explored in the discussion. Suffice to state that Slovenia (si) was the only region in which the importance of OAI-PMH endpoint was recognised and Cool URI management appeared evident.

**Table 4.** Summary counts and percentages of dead repositories by domain and OAI-PMH endpoint, categorized and ordered by two-letter country code (ISO 3166-1 alpha-2).

| Repository country | Total count | Dead count (domain) | Dead count (%) (domain) | Dead count (OAI-PMH) | Dead count (%) (OAI-PMH) | Domain / OAI-PMH differential | Domain / OAI-PMH differential (%) |
|---|---|---|---|---|---|---|---|
| ad | 1 | 0 | 0.0 | 0 | 0.0 | 0 | 0.0 |
| al | 3 | 0 | 0.0 | 0 | 0.0 | 0 | 0.0 |
| at | 78 | 5 | 6.4 | 22 | 28.2 | 17 | 21.8 |
| ba | 5 | 3 | 60.0 | 4 | 80.0 | 1 | 20.0 |
| be | 61 | 30 | 49.2 | 35 | 57.4 | 5 | 8.2 |
| bg | 21 | 10 | 47.6 | 11 | 52.4 | 1 | 4.8 |
| by | 61 | 12 | 19.7 | 29 | 47.5 | 17 | 27.9 |
| ch | 63 | 6 | 9.5 | 19 | 30.2 | 13 | 20.6 |
| cy | 11 | 2 | 18.2 | 5 | 45.5 | 3 | 27.3 |
| cz | 42 | 7 | 16.7 | 13 | 31.0 | 6 | 14.3 |
| de | 494 | 104 | 21.1 | 194 | 39.3 | 90 | 18.2 |
| dk | 30 | 10 | 33.3 | 14 | 46.7 | 4 | 13.3 |
| ee | 12 | 3 | 25.0 | 5 | 41.7 | 2 | 16.7 |
| es | 325 | 56 | 17.2 | 105 | 32.3 | 49 | 15.1 |
| fi | 43 | 13 | 30.2 | 18 | 41.9 | 5 | 11.6 |
| fr | 234 | 49 | 20.9 | 89 | 38.0 | 40 | 17.1 |
| gb | 528 | 125 | 23.7 | 215 | 40.7 | 90 | 17.0 |
| gr | 71 | 20 | 28.2 | 29 | 40.8 | 9 | 12.7 |
| hr | 185 | 13 | 7.0 | 31 | 16.8 | 18 | 9.7 |
| hu | 78 | 15 | 19.2 | 19 | 24.4 | 4 | 5.1 |
| ie | 56 | 9 | 16.1 | 23 | 41.1 | 14 | 25.0 |
| is | 5 | 2 | 40.0 | 3 | 60.0 | 1 | 20.0 |
| it | 211 | 62 | 29.4 | 96 | 45.5 | 34 | 16.1 |
| lt | 31 | 4 | 12.9 | 11 | 35.5 | 7 | 22.6 |
| lu | 5 | 4 | 80.0 | 4 | 80.0 | 0 | 0.0 |
| lv | 8 | 1 | 12.5 | 2 | 25.0 | 1 | 12.5 |
| md | 23 | 6 | 26.1 | 15 | 65.2 | 9 | 39.1 |
| mk | 8 | 2 | 25.0 | 4 | 50.0 | 2 | 25.0 |
| mt | 2 | 0 | 0.0 | 0 | 0.0 | 0 | 0.0 |
| nl | 105 | 22 | 21.0 | 32 | 30.5 | 10 | 9.5 |
| no | 92 | 21 | 22.8 | 24 | 26.1 | 3 | 3.3 |
| pl | 208 | 45 | 21.6 | 88 | 42.3 | 43 | 20.7 |
| pt | 102 | 26 | 25.5 | 32 | 31.4 | 6 | 5.9 |
| ro | 22 | 8 | 36.4 | 12 | 54.5 | 4 | 18.2 |
| rs | 99 | 8 | 8.1 | 12 | 12.1 | 4 | 4.0 |
| ru | 102 | 39 | 38.2 | 58 | 56.9 | 19 | 18.6 |
| se | 97 | 19 | 19.6 | 25 | 25.8 | 6 | 6.2 |
| si | 19 | 4 | 21.1 | 3 | 15.8 | -1 | -5.3 |
| sk | 9 | 2 | 22.2 | 5 | 55.6 | 3 | 33.3 |
| sm | 1 | 0 | 0.0 | 0 | 0.0 | 0 | 0.0 |
| ua | 200 | 57 | 28.5 | 97 | 48.5 | 40 | 20.0 |

Recall HTTP status request codes for every repository domain URL and its associated OAI-PMH endpoint were gathered. HTTP responses were logged against the repository entries within our registry dataset. These data reveal the nature of dead repository responses but also those which remain alive. They are set out in Table 5.

2320 repositories within our wider dataset of 3,751 returned a 200 response, with less returning a 200 for their OAI-PMH endpoint ($n$ = 1755). A not insignificant proportion of repositories returned redirection status codes (3XX) at both the domain ($n$ = 607) and endpoint levels ($n$ = 353). The significant number of 302 responses at the domain level is unusual ($n$ = 580) and worthy of comment in the discussion section. It can nevertheless be noted here that such a code indicates a requested resource has moved temporarily but will return to its original location later.

824 repositories were identified as being deceased, based on their domain. In Table 5 we see that DNS resolution errors (NXDOMAIN) were recorded in 751 cases, 91% of all negative responses. This NXDOMAIN count almost doubled for OAI-PMH endpoints ($n$ = 1400). A diversity of responses were logged across 5XX and 4XX, with the most common indicator of repository death after NXDOMAIN being a 404 response.

**Table 5.** HTTP response code counts against repository domains and OAI-PMH endpoints.

| Response code categories | HTTP response code | Repository count (domain) | Repository count (OAI-PMH) |
|---|---|---|---|
| **Non-existent domain** | NXDOMAIN | 751 | 1400 |
| **Server error responses (5XX)** | 504 | 1 | 0 |
| | 503 | 10 | 7 |
| | 502 | 3 | 1 |
| | 500 | 2 | 8 |
| **Client error responses (4XX)** | 422 | 0 | 4 |
| | 410 | 0 | 3 |
| | 404 | 38 | 106 |
| | 403 | 18 | 42 |
| | 400 | 1 | 67 |
| **Sub-total** | | 824 | 1638 |
| **Redirection status codes (3XX)** | 308 | 7 | 1 |
| | 307 | 13 | 10 |
| | 303 | 6 | 2 |
| | 302 | 580 | 339 |
| | 301 | 1 | 1 |
| **Sub-total** | | 607 | 353 |
| **Successful response** | 200 | 2320 | 1755 |
| **Total** | | 3751 | 3751 |

In separate but related analyses we can report that where a redirection was found at the domain level, a similar redirection to its corresponding OAI-PMH endpoint was not implemented. Instead, a total of 148 (Table 6) such endpoints were found to be dead, with NXDOMAIN the most common response for a repository domain reporting a 302 response. These examples accounted for 67% of all those identified; though it should also be noted that NXDOMAIN also arose from all the noted cases of 308, 307, 303, and 301. NXDOMAIN was therefore the response found in a total of 110 cases (74%). A mixture of client error responses (4XX) and server error responses comprised the small remainder. We can therefore report that close to one tenth (9%) of all unresponsive OAI-PMH endpoints were associated with a repository which redirected at the domain level but not at the endpoint level, a possible indicator that these may be zombie repositories.

**Table 6.** Recorded 3XX responses at repository domain level and corresponding responses at associated OAI-PMH endpoint.

| Redirection response (domain) | HTTP response (corresponding OAI-PMH endpoints) | Repository count (OAI-PMH) |
|---|---|---|
| 308 | NXDOMAIN | 5 |
| 307 | NXDOMAIN | 3 |
| 303 | NXDOMAIN | 2 |
| 302 | NXDOMAIN | 99 |
| 302 | 503 | 1 |
| 302 | 410 | 2 |
| 302 | 404 | 12 |
| 302 | 400 | 23 |
| 301 | NXDOMAIN | 1 |
| **Total** | | 148 |

## Date of Decease

### Wayback Machine Availability

Of the 824 repositories identified as dead within our dataset, it was possible to identify archived snapshots via the Wayback Machine Availability API for 680. The date of decease for these repositories begins in the year 2001 and continues until the present day; though, to control for repositories experiencing recent availability problems, we exclude snapshots captured after 2025-01-01.

To assess both the absolute growth and underlying dynamics of repository death, we employ both linear and logarithmic scaling. Linear and log-linear regression models were evaluated to characterize the growth of cumulative counts over time. Linear was used to model overall trends and support regression analysis, while logarithmic was applied to highlight growth phases and detect potential exponential behaviour.

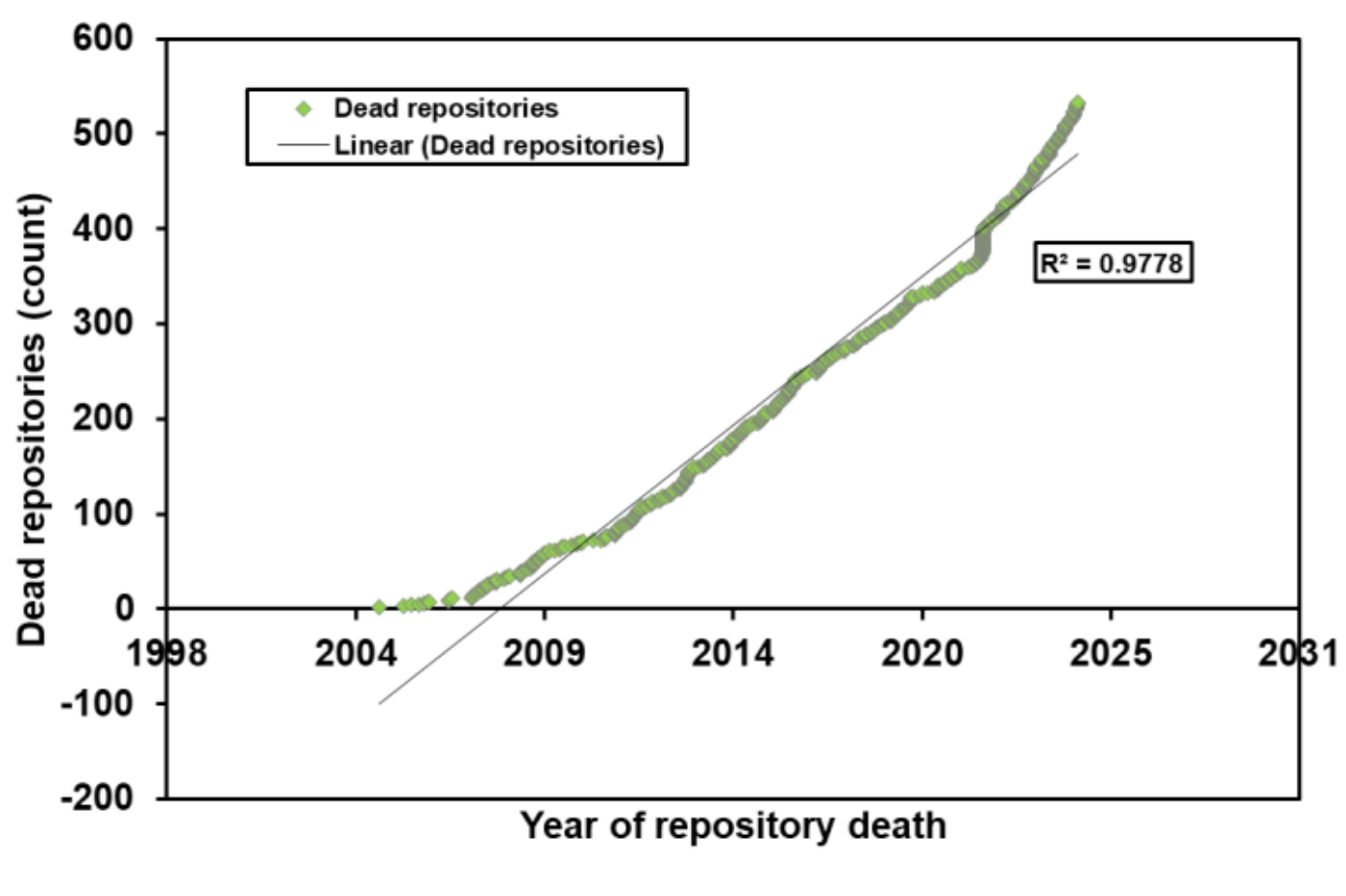

**Figure 2.** Growth of repository deaths over time, with linear regression included.

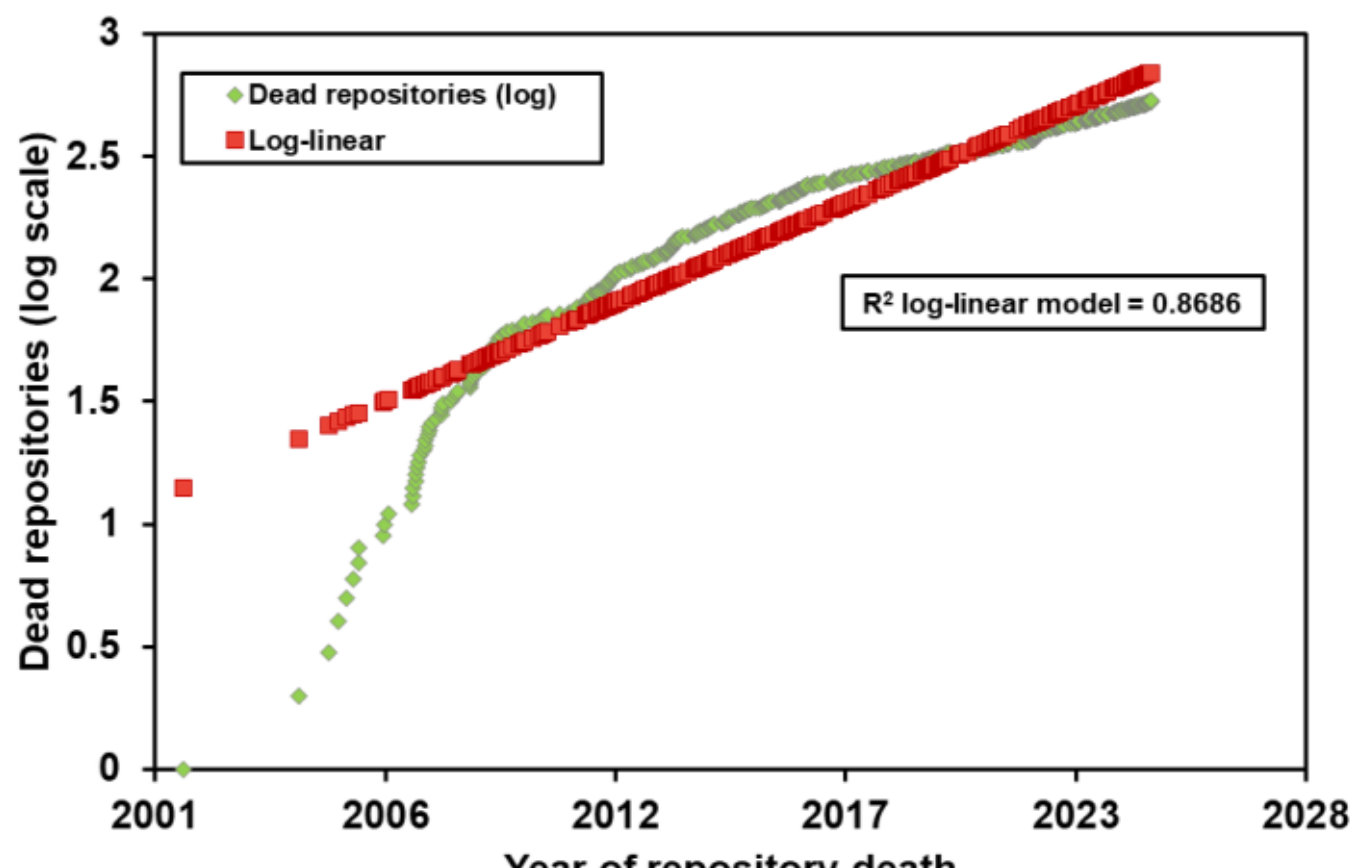

**Figure 3.** Growth of repository deaths over time (log), with log-linear regression included. Y-axis values represent the natural logarithm of the data in Figure 2.

We can observe from Figure2 the temporal profile of dates of decease in the linear model, in which a slight but steady acceleration in the cumulative total of dead repositories in the years following 2011 can be noted. Another period of acceleration occurs in 2022. The linear model demonstrated a superior fit ($R^2$ = .978) compared to the log-transformed model ($R^2$ = .869) in Figure 3, suggesting a strong linear association ($F(1, 531)$ = 21,049.13, $p < .001$). While the log-linear model revealed a statistically significant exponential component ($F(1, 531)$ = 3510.59, $p < .001$) and revealed interesting early-stage dynamics, the growth rate was modest (approximately 0.0202% per time unit), indicating that the cumulative increase was more consistent with additive rather than multiplicative growth. These findings support the use of a linear model for predictive purposes, though we should recall that this observation is based on a subset of Wayback Machine Availability data relating to our wider repository dataset.

## Rate of decay

We can also reconceptualize repository death as the rate of decay of once functioning, so-called 'mortal' repositories. We can do this by employing a classic decay function, which can be calculated where $y$ is the number of repositories remaining at the end of our temporal range, where $a$ is the number repositories at the beginning of that range, where $r$ is the rate of decay expressed as a decimal, and where $t$ equals the time elapsed which, in this case, is the number of years.

$$y = a(1 - r)^t$$

This yields an annual decay rate of 0.0107, meaning that over our 23-year period the number of functioning repositories shrinks by 1.07% every year. At this rate of decay, we can estimate that within a decade (i.e. by 2034) the number of functioning European repositories will shrink by a further 442, from the 2024 benchmark of 2,927.

Acknowledging that our data reveals deceased OAI-PMH endpoints are decaying more rapidly, we can similarly perform this analysis to better quantify OAI-PMH endpoint decay. Here the decay rate is 0.0237, indicating that death of OAI-PMH endpoints is far more pronounced at 2.47% per annum. We can therefore expect that the number of mortal OAI-PMH endpoints in 2034 will have declined by a further 681, from its 2024 benchmark of 2,108.

## Dead repository content in the scholarly literature

The final section of our results is concerned with mining the scholarly literature to measure the extent to which dead European repository content may have entered the scholarly record through reference or citation. Our mining seeks to identify citations and references as they are understood in scholarly communication (Nicolaisen, 2007). That is, an acknowledgement of 'intellectual debt' within academic text (i.e. citation), which refers the reader to details of a cited literature source (i.e. reference) within a list of works (Garfield, 1996). But they can also be understood as author *A* making 'reference' to author *B*'s work within academic text, through an in-text reference; while a 'citation' is when author *A* receives acknowledge from author *B* through an in-text citation (Albarrán & Ruiz-Castillo, 2011). Our mining seeks to identify where repository content may have been cited and/or referenced. It also identifies the citation of repository handles directly within academic text.

The results arising from mining the literature identified 12,040 unique scholarly works that cite repository content which we know through our analysis to be dead. These unique works generated a total of 19,248 references to dead content. This is because some works reference content in more than one dead repository, or multiple instances of scholarly content from the same dead repository. This is reflected in the associated measures of central tendency ($M = 1.6$, $Md = 1$, $Range = 1\text{-}70$, $IQR = 1$). Most scholarly works reference just one (> 67%) or two (> 20%) instances of dead repository content (Table 7) but a notable observation here are that references to dead repository content within unique works can often be multiple. Though our upper range is influenced by an extreme outlier (i.e. 70 references to dead content within a single work) – and though such double-digit examples do not skew the median, accounting for fewer than 42 references – we can count 1,614 examples of works citing or referring to ≥ 3 dead repository sources.

**Table 7.** Number of references from a single scholarly work to dead repository content and number of scholarly works within this assigned category.

| No. dead references within work | No. scholarly works with assigned number of dead references (count) | No. scholarly works with assigned number of dead references (%) |
|---|---|---|
| 1 | 8,075 | 67.07 |
| 2 | 2,351 | 19.53 |
| 3 | 906 | 7.52 |
| 4 | 354 | 2.94 |
| 5 | 177 | 1.47 |
| 6 | 79 | 0.66 |
| 7 | 37 | 0.31 |
| 8 | 14 | 0.12 |
| 9 | 5 | 0.04 |
| >9 | 42 | 0.35 |

An observation from our data are that dead repository content can remain alive in newly published literature for many years after the decease of a repository. Of the scholarly works derived from CORE within our dataset, 10,375 had a reliable date of publication which, combined with the established date of repository decease, could be used to calculate (in years) the extent to which dead content was or was not being cited in new literature. Key measures are summarized in Table 8.

**Table 8.** Summarization of dead and alive repository references in newly published literature, where an accurate date of publication available (from CORE). Negative values within the 'Years' category denotes the number of years after repository death that content was cited in published literature.

| | All references | 'Dead on arrival' repository references | 'Alive' repository references |
|---|---|---|---|
| **Total count** | 10,375 | 2,801 | 7,575 |
| **Total %** | 100.00 | 36.98 | 73.01 |
| | *Years (+/-)* | *Years (+/-)* | *Years (+/-)* |
| **Mean (*M*)** | 3.01 | -3.57 | 5.44 |
| **Median (*Md*)** | 3.00 | -2.00 | 5.00 |
| **Standard deviation (*SD*)** | 5.50 | 3.78 | 3.77 |

Most scholarly works (circa 73%) referenced or cited repository content that was alive when the work was published. However, we found that almost 37% ($n$ = 2,801) of these works cited repository content that was already dead upon its publication ('dead on arrival'). This repository content was found to be already dead, on average, for > 3.5 years when the scholarly work was published ($M$ = -3.57, $Md$ = -2.00, $SD$ = 3.78). Results also appear to identify instances of dead repository content being cited in new literature as many as 7 years, 10 years, and sometimes even 17 years after a repository was known to be dead. This suggests that the consequence of repository death can linger in the scholarly record, with authors continuing to cite dead content in new literature long after that content no longer exists. This phenomenon is concerning. We proffer explanations for it in the discussion section.

The spread of 'dead on arrival' references is better appreciated in Figure 4, which illustrates the year of repository decease and the 'distance' (in number of years) between publication of the scholarly work and the year of repository death. This distance may be positive or negative. Positive values on the Y-axis denote the number of years the repository was alive after its citation within the scholarly works. Conversely, negative values denote the number and extent of 'dead on arrival' references within published literature.

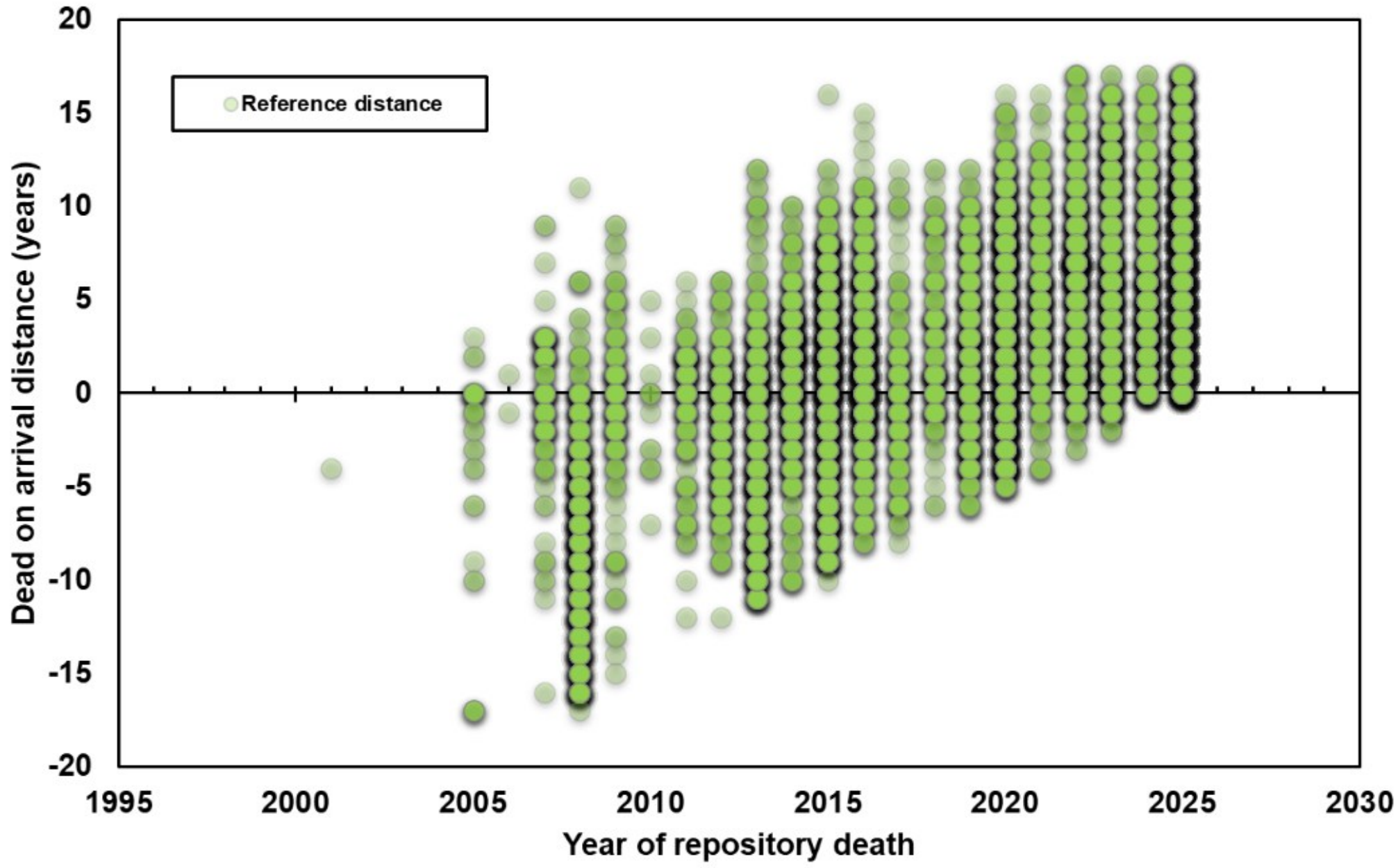

**Figure 4.** Dead and alive references to repository content within scholarly published literature. Figure highlights 'dead on arrival' references. Negative values on Y axis indicate level to which repository was already dead (in years) when cited in scholarly literature.

# Discussion

We have attempted to measure repository persistence and the impact impersistance has on the scholarly record. The results make a series of important observations. In our discussion we will comment on some of the most significant.

The study indicates that over a fifth of the repositories within our dataset are dead, with institutional repositories forming a significant proportion of these deaths (>75%). The machine interfaces of repositories within the dataset were also found to be dead in more than a third of cases. Again, institutional repositories were highly represented, accounting for almost 80% of all dead OAI-PMH endpoints; though we should remember that some repository deaths were found to be proportionally large in some category types (e.g. governmental). These findings corroborate emerging concerns that scholarly repositories are often mismanaged, despite their importance to scholarship and open research. This mismanagement unfortunately contributes to the persistence crisis and raises questions about the trustworthiness of the custodian organizations of such scholarly resources. This applies particularly to institutional repositories, a dominant organizational manager of which are research libraries.

The nature of the study limits our ability to comment on the organizational reasons for why certain organizations have precipitated repository death. However, the multitude of NXDOMAIN HTTP responses for repository domains and their OAI-PMH endpoints, as well as 4XX and 5XX responses, suggests these repositories have been neglected and poorly maintained, with termination ultimately considered a preferable course of action or indeed where no course of action has been considered at all. Rothfritz et al. (2025) and Bamgbose et al. (2025) highlight the sustainability pressures threatening the operation of repositories, which thereby undermine their trustworthiness. This includes financial constraints and a level of organizational disengagement from repository operations, resulting in low awareness of the consequences arising from repository death. The volume of dead repositories identified through this study is such that we must therefore conclude a lack of understanding exists in many organizations, about the core function of a repository, the implied commitment to long-term access, and the scholarly consequences of shutting one down. Concern about the technical capacity, digital skills, and knowledge transfer within repository responsible teams, partly related to financial constraints, is likely a contributing factor. Research draws attention to an emerging skills deficit within research libraries, digital archives, and digital libraries, which would impact effective institutional repository management (Cope & Baker, 2017; Recker et al., 2024; Tait et al., 2016), including the 'platformization' of repositories within some research libraries (Plantin & Thomer, 2025). Outsourcing of technical infrastructure has also been found to usurp digital capacity (SCONUL, 2025), limiting the ability of some organizations to respond responsibly to long-term repository management. Organizations or groups responsible for disciplinary repositories may face separate challenges. These can include governance difficulties, a lack of funding, and disinterest from communities of practice (Björk, 2014; Rieger, 2012).

A notable additional factor which is likely contributing to the growth of dead institutional repositories will be the increased deployment of current research information systems (CRIS) at European universities (Biesenbender et al., 2019; Bryant et al., 2018). Though CRIS software is designed to support institutional research monitoring (Fabre et al., 2021) and offers a system purpose and function that is distinct from repositories, many institutions have been observed merging their functions for organizational convenience (de Castro et al., 2014). In a global survey of CRIS deployment at research organizations, Bryant et al. (2018) confirmed a particular intensity for such software within Europe but also a trend towards using the software as a repository substitute, with CRIS software functioning as the 'default' repository for scholarly literature in > 50% of respondents. Schöpfel and Azeroual (2021), Schöpfel et al. (2024), and Fernandes and Pinto (2018) all report on the increased convergence between repositories and CRIS software and the resulting management challenges. Schöpfel et al. (2024) note too that this shift may not be to the benefit of open research and 'underscores' the critical challenge arising from such a transformation.

This so-called 'CRIS-as-IR' trend is worrisome because it tends to result in the death of a repository. It also places primacy upon software that is rarely open source, is often ill-suited to long-term digital object management, and rarely displays adequate support for Cool URI management. This therefore increases the likelihood that the future location of scholarly resources may again be disrupted, further exacerbating the diffusion of impersistent references in the scholarly literature. At time of writing it can be noted from the Directory of Research Information Systems (DRIS)[10] (maintained by EuroCRIS) that Europe is among the strongest regions for adopting CRIS solutions, with the United Kingdom, Italy, Spain, Germany, and Poland particularly strong. This may go some way to explaining why some of these countries were also among the regions to host the largest concentration of dead repositories.

Our analysis of HTTP responses elicited indications that some live repositories were in fact zombie repositories. Recall that zombie repositories are those that are killed but then 'reanimated' under a new domain, often using different software and serving content from alternative locations. We can observe indications of this in the implementation of some 3XX redirections, where URI management (via redirection) at the domain level is displayed but not at the OAI-PMH endpoint level, where NXDOMAIN, 404s, etc are instead recorded. Since the OAI-PMH endpoints of these zombie repositories are dead – and ergo the OAI identifiers associated with individual repository resources too – it is probable that limited URI management will have been performed on the corresponding URI context or accession identifiers for individual repository resources. In these instances, managing organizations appear to have an awareness that users and machines will be disrupted by creating a zombie repository and so provide a redirection at the domain level. It can be presumed that this action is taken to redirect users who have bookmarked the repository location, or to avoid the breaking of domain-level hyperlinks elsewhere on the web. But either through a lack of migration planning or technical limitations, they do not seek to ensure URI persistence beyond the domain level. It is for these reasons that Berners-Lee's original contention was that organizations often displayed a "lack of forethought" in their URI management (Berners-Lee, 1998).

Findings arising from our analysis of repository date of decease and the rate of repository decay suggest additive growth in European repository deaths is occurring. Based on our data, there are indications that dead repositories emerge more quickly than new repositories are established, at least within Europe. As the region displaying the highest concentration of registered repositories, Europe is likely to considerably erode the total number of active repositories globally. The predictive potential of these observations must always be weighed against the prospect that new, more persistent repositories may be launched in future. We must also consider the possibility that, were new repositories to be launched, they may simply perpetuate repository death given repository mismanagement we have observed. Moreover, we must consider the possibility that a proportion of any 'new' European repositories may be zombie repositories, reanimated following prior repository death. An exact understanding of the wider impact of this observation of decay is therefore difficult to establish. Reference data from ROAR on recorded repository registrations (globally) since 1991 indicates that the launching of new repositories peaked in 2012, with a steady decline thereafter. Zenodo[11] was launched in 2013 as a repository solution to capture the 'long tail' of European research work (Amorim et al., 2015). A consequence may have been to usurp the launching of new repositories in the years since 2012. Zenodo fulfils a wide variety of repository use cases that hitherto would have necessitated the creation of a dedicated repository and its impact on scholarly communication more generally has been vast (Crespo Garrido et al., 2025).

Recall from our data collection that the parsing of HTTP 301 responses associated with HTTP to HTTPS redirects to a cognate domain were treated as a 200 response. This process surfaced frequent inappropriate use of the HTTP 302 response by repositories when a 301 response was expected. HTTP 302 indicates that a requested resource has moved temporarily but will return to its original location later (Fielding et al., 2022). For this reason, software agents, such as a search engine bot, will typically retain an original resource location and elect not to update its indexes with a location it has been informed is temporary (Mozilla Foundation, 2025). It is notable that within our dataset, 580 repositories returned a 302 response instead of an expected 301 response for what appeared to be a permanent HTTP to HTTPS redirect. Usually this was an HTTP to HTTPS redirect at the root domain, though some appeared permanent redirects to new unrelated locations. This observation suggests that many repositories are mistakenly configured to serve a 302 response when they should in fact be serving a 301. Such local misconfiguration of repository infrastructure can negatively influence search engine discovery and is something for the repository community to act upon.

[10] Directory of Research Information Systems (DRIS): https://eurocris.org/services/dris

[11] Zenodo: https://zenodo.org/

Understanding the extent of European repository impersistence was the first step to measuring the impact it might be having on the scholarly record. We found that it was possible to surface almost 20,000 references to dead repository content through our mining of the scholarly literature. That these dead references arise from circa 12,000 unique scholarly works reveals the proclivity some authors display in citing multiple repository resources within a single scholarly work. Unfortunately, it also reveals a level of citation confidence in repositories which should exist but does not. Instead, organizational mismanagement of repository infrastructure has undermined the scholarly record and contributed to the reference rot crisis. Despite its size and scope, CORE is not an exhaustive corpus of scholarly literature, nor are all scholarly resources available for TDM interrogation within CORE. On this basis we can reliably assume that the number of references or citations to dead repository resources within the literature is likely higher than reported here. Similarly, there is likely a higher proportion of 'dead on arrival' references within the literature too.

A notable finding emerging from our study are that the aftereffects of repository death can be found in the newly emerging scholarly record, years after a repository is known to have died. So-called 'dead on arrival' references were therefore an interesting but concerning observation given the volume of such references detected ($n$ = 2,801).

One possible explanation for this might be the increased use of reference management software and the growth of personal information management (PIM) among scholars. Scholars producing systematic reviews have relied on reference management software for some time (Lorenzetti & Ghali, 2013), and key review guides in this space, such as the 'Cochrane Handbook for Systematic Review of Interventions', even propose their own software solutions for this purpose (Higgins et al., 2024). Software solutions not only support the capture and management of metadata describing found resources, but also the caching, saving, and annotation of associated PDF full-text documents, such as Zotero, Mendeley, EndNote, etc. (Speare, 2018; Williams & Woods, 2024). More recent studies indicate that the ability to "save and organize PDF files" within solutions is an essential feature for many users (Nitsos et al., 2022). It is therefore conceivable that an overreliance on such solutions by authors has emerged in which authors continue to re-use dead repository content within the writing process without realising it is dead. This is because authors may only be referring to what has been previously captured, or annotated locally or in the cloud, rather than verifying the reachability of the resource they are using during writing and citing. Some solutions support features that can mitigate link rot. For example, Zotero's support for application plugins has enabled the creation of 'Zotero Memento' (leonkt, 2019/2021), which observes the Memento Protocol (Jones, Klein, Sompel, et al., 2021). However, only a minority of scholars are likely to use these features since usage is predicated on understanding the issue of reference rot in the first place. The issue is that scholars generally expect repositories to be persistent and that the scholarly resources they have located in the past will remain available. This is, after all, the way in which repositories have been advocated to scholarly users since the beginning (Lynch, 2003). The death or zombification of a repository therefore further exacerbates the impact of repository impersistence to open research and scholarship more generally.

It could be suggested that the emergence of certification initiatives, such as CoreTrustSeal (CoreTrustSeal Standards and Certification Board, 2022), as well as projects designed to stimulate maturity in FAIR-enabling repositories (van Lieshout et al., 2025), is tacit sectoral acknowledgement of a problem that has hitherto been unquantifiable. Without adherence to trustworthy repository frameworks, research performing organizations may always fail to acknowledge the long-term commitment arising from repository management. They may also lack cognisance of possible consequences to scholarship, users, or open scholarly infrastructure. Efforts by the European Open Science Cloud (EOSC) (Burgelman, 2021) to grow a resilient, open, and 'trusted' environment in which scholars can easily publish, locate and reuse scholarly content has identified a need to improve FAIRness and trustworthiness in repository infrastructure. Indeed, there is recognition that trustworthy digital repositories are central to the realisation of the EOSC (Directorate-General for Research and Innovation (European Commission), 2018). Projects emerging from the EOSC, such as FAIR-IMPACT (Dillo et al., 2024), concluded that greater transparency in repository processes was necessary to increase their overall trustworthiness (Grootveld et al., 2025). Subsequent similar work under the auspices of the FIDELIS project has proposed resources, including the Transparent Trustworthy Repository Attributes Matrix (TTRAM) which is designed to serve as a reference model and can assist in assessing the extent to which repository operations around digital object management, organizational infrastructure, and technology contribute to scholarly transparency and trustworthiness (L'Hours et al., 2025).

# Limitations and future research

Our study has limitations, some of which may motivate further research. We noted in our methods that not all scholarly works in CORE have full text available for TDM interrogation, nor does CORE provide an inventory of all known literature. This means that an authoritative, universal picture of total dead repository content diffusion in the scholarly literature will never be possible. However, CORE remains the largest such open dataset available and therefore a useful indicator. Future work could explore similar analyses while enlarging the TDM dataset to include other sources. An obvious expansion of the study would also be to expand analyses beyond Europe to include all available repositories. Quantifying the global challenge of repository impersistence to scholarship would be possible, as would the observation of geographic or continental differences.

In relation to 'dead on arrival' references, we should acknowledge the limitation that CORE's aggregation of scholarly literature can span many different work types, e.g. accepted manuscript, preprint, version of record, etc. It is therefore conceivable that a subset of the identified 'dead on arrival' references within, say, a preprint or accepted manuscript, were later identified and corrected during editorial steps and/or typesetting by a publisher within a Version of Record. This is difficult to detect and ergo quantify without significant additional analysis. It is therefore a suggestion for future research.

A final limitation worthy of noting pertains to the HTTP GET approach employed to capture repository response codes. It should be acknowledged that while it was advantageous in our context to use a distributed cloud infrastructure for this purpose, it is conceivable that requests from specific cloud IP ranges may have been treated differently to fixed client requests. We surmise that the influence of such variability was minimal in this instance because those services displaying erratic availability and/or errors were re-tested and spot-checked.

The findings of this study could be used to define recommendations and/or guidance that can support organisations and the repositories they host to develop and implement policies and practices that ensure greater sustainability and limit the number of dead repositories emerging in the years ahead.

Finally, a productive area for future research would be to better understand the way in which dead repository content has been cited or referenced in the scholarly literature. Circa 20,000 references to dead repository content were found to have entered the scholarly record as part of this study. But it has long been known that not all references or citations are equal (Moravcsik & Murugesan, 1975). This is because scholarly works are cited by authors in a variety of ways and for a variety of different purposes, e.g. is the reference 'organic' or 'perfunctory', or, 'evolutionary' or 'juxtapositional'? Even the location of a citation within a scholarly work can be a predictor of its academic utility (Cano, 1989). This means that the negative scholarly impact arising from dead repository content will vary depending on how it has been referenced or cited by the author. A more granular quantification of impact is therefore possible by analysing, categorizing, and measuring the way in which dead repository content has been cited within the works themselves. Though computationally intensive to perform, the use of research methods harnessing sentiment analysis may help to reveal how dead repository content has been cited, as well as provide a more nuanced understanding of its relative negative impact.

# Conclusion

Open repositories have become essential components of open scholarly infrastructure in recent decades. They provide infrastructure to support the storage, discovery, and impact of research. The persistence of this infrastructure is therefore critical to scholarship, supporting key aspects of the open research agenda. Instead of constituting reliable nodes in open scholarly infrastructure, this study has exposed concerning weaknesses in the persistence of European repository infrastructure, with significant levels of repository impersistence detected. These examples of impersistence were also found to have compromised the scholarly record in thousands of cases (including within 'dead on arrival' references), thereby hindering content discovery, research citation, verification, and reproducibility. In other words, infrastructure that has been designed and managed by organizations to better support the goals of open scholarship are in many cases damaging it.

It is clear from our work that many organizations operating repositories experience challenges managing them, including the misconfiguration of HTTP redirects, poor URI management, and a technical naivety in the strategic management of repositories such that the death or zombification of otherwise healthy repositories arises. The causes for this predicament are complex, as described earlier. As the findings of this study highlight, the consequences of repository mismanagement to scholarship are significant. These consequences need to be better acknowledged and understood by managing organizations. This is particularly relevant to university research libraries which, as part of this study, were found to have hosted a disproportionate number of found dead repositories, and which we have noted generally purport to be leaders in open research policy and practice.

# Data Availability Statement

All data and software code underpinning the research documented in this article are available at: https://doi.org/10.5525/gla.researchdata.2101

# Acknowledgements

We wish to acknowledge insightful online discussions about repository persistence with Paul Walk (Antleaf), Michael P. Taylor (Index Data and University of Bristol), and Herbert Van de Sompel (DANS and University of Ghent).

Joy Davidson acknowledges funding from the European Union's Horizon Europe Framework Programme for Research and Innovation, under grant agreement 101188078. FIDELIS: Establishing a European Network of Trustworthy Digital Repositories (https://doi.org/10.3030/101188078).

# References

Albarrán, P., & Ruiz-Castillo. (2011). References made and citations received by scientific articles. *Journal of the American Society for Information Science and Technology*, 62(1), 40-49. https://doi.org/10.1002/asi.21448

Amorim, R. C., Castro, J. A., da Silva, J. R., & Ribeiro, C. (2015). A Comparative Study of Platforms for Research Data Management: Interoperability, Metadata Capabilities and Integration Potential. In A. Rocha, A. M. Correia, S. Costanzo, & L. P. Reis (Eds), *New Contributions in Information Systems and Technologies* (Vol. 353, pp. 101–111). Springer International Publishing. https://doi.org/10.1007/978-3-319-16486-1_10

Baglioni, M., Pavone, G., Mannocci, A., & Manghi, P. (2025). Towards the interoperability of scholarly repository registries. *International Journal on Digital Libraries*, 26(1), 2. https://doi.org/10.1007/s00799-025-00414-y

Baker, M. (2016). 1,500 scientists lift the lid on reproducibility. *Nature*, 533(7604), 452–454. https://doi.org/10.1038/533452a

Bamgbose, A. A., Mohd, M., Tengku Wook, T. S. M., & Mohamed, H. (2025). Are digital repositories for university libraries trustworthy?: A systematic review. *Business Information Review*, 02663821251328830. https://doi.org/10.1177/02663821251328830

Berners-Lee, T. (1998). *Cool URIs don't change*. W3C. https://www.w3.org/Provider/Style/URI

Biesenbender, S., Petersohn, S., & Thiedig, C. (2019). Using Current Research Information Systems (CRIS) to showcase national and institutional research (potential): Research information systems in the context of Open Science. *Procedia Computer Science*, 146, 142–155. https://doi.org/10.1016/j.procs.2019.01.089

Björk, B.-C. (2014). Open access subject repositories: An overview. *Journal of the Association for Information Science and Technology*, 65(4), 698–706. https://doi.org/10.1002/asi.23021

Brembs, B. (2018). Prestigious Science Journals Struggle to Reach Even Average Reliability. *Frontiers in Human Neuroscience*, 12. https://doi.org/10.3389/fnhum.2018.00037

Bryant, R., Clements, A., de Castro, P., Cantrell, J., Dortmund, A., Fransen, J., Gallagher, P., & Mennielli, M. (2018). *Practices and Patterns in Research Information Management: Findings from a Global Survey*. OCLC Research. https://doi.org/10.25333/BGFG-D241

Burgelman, J.-C. (2021). Politics and Open Science: How the European Open Science Cloud Became Reality (the Untold Story). *Data Intelligence*, 3(1), 5–19. https://doi.org/10.1162/dint_a_00069

Burns, C. S., Lana, A., & Budd, J. M. (2013). Institutional Repositories: Exploration of Costs and Value. *D-Lib Magazine*, 19(1/2). https://doi.org/10.1045/january2013-burns

Cano, V. (1989). Citation behavior: Classification, utility, and location. *Journal of the American Society for Information Science*, 40(4), 284–290. https://doi.org/10.1002/(SICI)1097-4571(198907)40:4%253C284::AID-ASI10%253E3.0.CO;2-Z

Carnevale, R. J., & Aronsky, D. (2007). The life and death of URLs in five biomedical informatics journals. *International Journal of Medical Informatics*, 76(4), 269–273. https://doi.org/10.1016/j.ijmedinf.2005.12.001

Collins, S., Genova, F., Harrower, N., Hodson, S., Jones, S., Laaksonen, L., Mietchen, D., Petrauskaité, R., & Wittenburg, P. (2018). *Turning FAIR into reality: Final report and action plan from the European Commission expert group on FAIR data*. Publications Office of the European Union. https://doi.org/10.2777/1524

Confederation of Open Access Repositories. (2022). *COAR Community Framework for Good Practices in Repositories, Version 2*. Zenodo. https://doi.org/10.5281/ZENODO.7108101

Cope, J., & Baker, J. (2017). Library Carpentry: Software Skills Training for Library Professionals. *International Journal of Digital Curation*, 12(2), 266–273. https://doi.org/10.2218/ijdc.v12i2.576

CoreTrustSeal Standards and Certification Board. (2022). *CoreTrustSeal Requirements 2023-2025*. Zenodo. https://doi.org/10.5281/zenodo.7051012

Crespo Garrido, I. del R., Loureiro García, M., & Gutleber, J. (2025). The Value of an Open Scientific Data and Documentation Platform in a Global Project: The Case of Zenodo. In J. Gutleber & P. Charitos (Eds), *The Economics of Big Science 2.0: Essays by Leading Scientists and Policymakers* (pp. 181–200). Springer Nature Switzerland. https://doi.org/10.1007/978-3-031-60931-2_14

Crow, R. (2002). *The Case for Institutional Repositories: A SPARC Position Paper*. SPARC. https://web.archive.org/web/20190307175704id_/http://pdfs.semanticscholar.org/da92/25f01aced7586efb7ea8f82406371374b06f.pdf

DANS., DCC., & EFIS. (2022). *European Research Data Landscape: Final report*. Publications Office of the European Union. https://doi.org/10.2777/3648

de Castro, P., Shearer, K., & Summann, F. (2014). The Gradual Merging of Repository and CRIS Solutions to Meet Institutional Research Information Management Requirements. *Procedia Computer Science*, 33, 39–46. https://doi.org/10.1016/j.procs.2014.06.007

Dillo, I., Ulrich, R., Huber, R., L'Hours, H., Davidson, J., Neidiger, C., Parkes, O., Everhardt, M., Reijnhoudt, L., Meijas, G., Ramezani, P., Verburg, M., & Priddy, M. (2024, August 29). Exposing repository information to foster connections and trust: Evaluating and implementing guidelines [Poster]. *International Conference on Digital Preservation (iPRES2024)*, Ghent, Belgium. https://doi.org/10.5281/ZENODO.13495834

Dimitrova, D., & Bugeja, M. (2007). Raising the Dead: Recovery of Decayed On-line Citations. *American Communication Journal*, 9(2). http://ac-journal.org/journal/2007/Summer/2RaisingtheDead.pdf

Directorate-General for Research and Innovation (European Commission). (2018). *Turning FAIR into reality: Final report and action plan from the European Commission expert group on FAIR data*. Publications Office of the European Union. https://doi.org/10.2777/1524

Dong, D., & Tay, C. H. A. (2023). Discoverability and Search Engine Visibility of Repository Platforms. In L. Woolcott & A. Shiri, *Discoverability in Digital Repositories* (1st edn, pp. 157–175). Routledge. https://doi.org/10.4324/9781003216438-11

Ducut, E., Liu, F., & Fontelo, P. (2008). An update on Uniform Resource Locator (URL) decay in MEDLINE abstracts and measures for its mitigation. *BMC Medical Informatics and Decision Making*, 8(1), 23. https://doi.org/10.1186/1472-6947-8-23

Eysenbach, G., & Trudel, M. (2005). Going, Going, Still There: Using the WebCite Service to Permanently Archive Cited Web Pages. *Journal of Medical Internet Research*, 7(5), e920. https://doi.org/10.2196/jmir.7.5.e60

Fabre, R., Egret, D., Schöpfel, J., & Azeroual, O. (2021). Evaluating the scientific impact of research infrastructures: The role of current research information systems. *Quantitative Science Studies*, 2(1), 42–64. https://doi.org/10.1162/qss_a_00111

Fernades, S., & Pinto, M. J. (2018). From the Institutional Repository to a CRIS system: what challenges? *Qualitative and Quantitative Methods in Libraries*, 7(3), 431-487. https://www.qqml-journal.net/index.php/qqml/article/view/494

Fielding, R. T., Nottingham, M., & Reschke, J. (2022). *HTTP Semantics [RFC 9110] (Request for Comments No. RFC 9110)*. Internet Engineering Task Force. https://doi.org/10.17487/RFC9110

Francke, H., Gamalielsson, J., & Lundell, B. (2017). Institutional repositories as infrastructures for long-term preservation. *Information Research*, 22(2). https://informationr.net/ir/22-2/paper757.html

Garfield, E. (1996). When to cite. *The Library Quarterly*, 66(4), 449-458. https://doi.org/10.1086/602912

Gertler, A. L., & Bullock, J. G. (2017). Reference Rot: An Emerging Threat to Transparency in Political Science. *Political Science and Politics*, 50(1), 166–171. https://doi.org/10.1017/S1049096516002353

Grootveld, M., Fink, A. S., Jonquet, C., González Guardia, E., Dillo, I., Nordling, J., Davidson, J., Marjamaa-Mankinen, L., Verburg, M., Priddy, M., GRAU, N., & Pittonet Gaiarin, S. (2025). *D1.3 Recommendations for a FAIR EOSC - White Paper of the FAIR-IMPACT Synchronisation Force (Version 1.0)*. Zenodo. https://doi.org/10.5281/zenodo.14979704

Helgesson, G., & Bülow, W. (2023). Research Integrity and Hidden Value Conflicts. *Journal of Academic Ethics*, 21(1), 113–123. https://doi.org/10.1007/s10805-021-09442-0

Higgins, J., Thomas, J., Chandler, J., Cumpston, M., Li, T., Page, M., & Welch, V. (Eds). (2024). *Cochrane Handbook for Systematic Reviews of Interventions version 6.5*. Cochrane. https://www.cochrane.org/authors/handbooks-and-manuals/handbook/current

Hockx-Yu, H. (2006). Digital preservation in the context of institutional repositories. *Program*, 40(3), 232–243. https://doi.org/10.1108/00330330610681312

Jefferies, N., Freeman, A., Notay, B., Knoth, P., Pentz, E., Buys, M., & Stewart, S. (2022). Open scholarship infrastructures and sustainability. *Oxford Festival of Open Scholarship 2022*, 7-18 March 2022, [Oxford]. Bodleian Libraries, University of Oxford. https://ora.ox.ac.uk/objects/uuid:7108677d-3092-47ff-b749-3479e4be0020

Jones, S. M., Klein, M., Sompel, H. V. de, Nelson, M. L., & Weigle, M. C. (2021). Interoperability for Accessing Versions of Web Resources with the Memento Protocol. In D. Gomes, E. Demidova, J. Winters, & T. Risse (Eds), *The Past Web: Exploring Web Archives* (pp. 101–126). Springer International Publishing. https://doi.org/10.1007/978-3-030-63291-5_9

Jones, S. M., Klein, M., & Van de Sompel, H. (2021). Robustifying Links To Combat Reference Rot. *Code4Lib Journal*, 50. https://journal.code4lib.org/articles/15509

Jones, S. M., Sompel, H. V. de, Shankar, H., Klein, M., Tobin, R., & Grover, C. (2016). Scholarly Context Adrift: Three out of Four URI References Lead to Changed Content. *PLOS ONE,* 11(12), e0167475. https://doi.org/10.1371/journal.pone.0167475

Jouneau, T., Verburg, M., Horton, L., van Geest, G., Tujunen, S., Kallio, J., Paulsen, T., Duvaud, S., Recker, J., Holthe-Tveit, Å. J., Thorpe, D. E., Conzett, P., Kleemola, M., Kalaitzi, V., Huber, R., Jonquet, C., Aguilar, F., Forshaug, A. K., Alaterä, T. J., & Esteves, E. (2025). *FIDELIS landscape survey analysis (Version 1.0)*. Zenodo. https://doi.org/10.5281/zenodo.15744996

Kelly, E. J. (2023). Discoverability Beyond the Library. In L. Woolcott & A. Shiri, *Discoverability in Digital Repositories* (1st edn, pp. 137–156). Routledge. https://doi.org/10.4324/9781003216438-10

Klein, M., & Balakireva, L. (2022). An extended analysis of the persistence of persistent identifiers of the scholarly web. *International Journal on Digital Libraries*, 23, 5–17. https://doi.org/10.1007/s00799-021-00315-w

Klein, M., Balakireva, L., & Shankar, H. (2019). Evaluating Memento Service Optimizations. *2019 ACM/IEEE Joint Conference on Digital Libraries (JCDL)*, 182–185. https://doi.org/10.1109/JCDL.2019.00034

Klein, M., Shankar, H., Balakireva, L., & Sompel, H. V. de. (2019). The Memento Tracer Framework: Balancing Quality and Scalability for Web Archiving. *International Conference on Theory and Practice of Digital Libraries.* (Vol. 11799, pp. 163–176). https://doi.org/10.1007/978-3-030-30760-8_15

Klein, M., Shankar, H., & Van de Sompel, H. (2018). Robust Links in Scholarly Communication. *Proceedings of the 18th ACM/IEEE on Joint Conference on Digital Libraries*, 357–358. https://doi.org/10.1145/3197026.3203885

Klein, M., Sompel, H. V. de, Sanderson, R., Shankar, H., Balakireva, L., Zhou, K., & Tobin, R. (2014). Scholarly Context Not Found: One in Five Articles Suffers from Reference Rot. *PLOS ONE,* 9(12), e115253. https://doi.org/10.1371/journal.pone.0115253

Knoth, P. (2025, July 3). Managing Machine Access to Open Repositories in the Age of Generative AI. *The 20th International Conference on Open Repositories*, Chicago, IL. https://doi.org/10.5281/ZENODO.15794663

Knoth, P., Herrmannova, D., Cancellieri, M., Anastasiou, L., Pontika, N., Pearce, S., Gyawali, B., & Pride, D. (2023). CORE: A Global Aggregation Service for Open Access Papers. *Scientific Data*, 10(1), 366. https://doi.org/10.1038/s41597-023-02208-w

leonkt. (2021). *Leonkt/zotero-memento (Version 1.1.1)* [JavaScript]. https://github.com/leonkt/zotero-memento (Original work published 2019)

L'Hours, H., Kleemola, M., Parkes, O., Recker, J., Duvaud, S., van Horik, R., Alaterä, T. J., Liberante, F., Conzett, P., Kaartinen, H., Bäckman, S., & Esteves, E. (2025). *FIDELIS TTRAMatrix v01.00 Introduction and Overview*. Zenodo. https://doi.org/10.5281/zenodo.17144159

Lorenzetti, D. L., & Ghali, W. A. (2013). Reference management software for systematic reviews and meta-analyses: An exploration of usage and usability. *BMC Medical Research Methodology*, 13(1), 141. https://doi.org/10.1186/1471-2288-13-141

Lynch, C. A. (2002). *Preserving Digital Information to Support Scholarship* (M. Devlin, R. C. Larson, & J. W. Meyerson, Eds). EDUCAUSE. https://www.cni.org/wp-content/uploads/2002/08/preserving-digital-info.pdf

Lynch, C. A. (2003). Institutional Repositories: Essential Infrastructure For Scholarship In *The Digital Age. Portal: Libraries and the Academy*, 3(2), 327–336. https://doi.org/10.1353/pla.2003.0039

Mabe, A., Nelson, M. L., & Weigle, M. C. (2021). Extending Chromium: Memento-Aware Browser. *2021 ACM/IEEE Joint Conference on Digital Libraries (JCDL)*, 310–311. https://doi.org/10.1109/JCDL52503.2021.00046

Mabe, A., Nelson, M. L., & Weigle, M. C. (2022). A Chromium-Based Memento-Aware Web Browser. In G. Silvello, O. Corcho, P. Manghi, G. M. Di Nunzio, K. Golub, N. Ferro, & A. Poggi (Eds), *Linking Theory and Practice of Digital Libraries* (pp. 147–160). Springer International Publishing. https://doi.org/10.1007/978-3-031-16802-4_12

Macgregor, G. (2023). Digital repositories and discoverability: Definitions and typology. In L. Woolcott & A. Shiri, *Discoverability in Digital Repositories* (pp. 11–31). Routledge. https://doi.org/10.4324/9781003216438-3

Macgregor, G. (2025, April 2). Revisiting the importance of persistence in the scholarly web. *THE Digital Universities UK Conference,* Lancaster University. https://eprints.gla.ac.uk/352395/

Macgregor, G., Lancho-Barrantes, B. S., & Pennington, D. R. (2023). Measuring the concept of PID literacy: User perceptions and understanding of PIDs in support of open scholarly infrastructure. *Open Information Science*, 7(1). https://doi.org/10.1515/opis-2022-0142

Mahanama, B., Balakireva, L., Jayarathna, S., Nelson, M., & Klein, M. (2022). Memento validator: A toolset for memento compliance testing. *Proceedings of the 22nd ACM/IEEE Joint Conference on Digital Libraries*, 1–3. https://doi.org/10.1145/3529372.3533297

Martin-Segura, S., Lopez-Pellicer, F. J., Nogueras-Iso, J., Lacasta, J., & Zarazaga-Soria, F. J. (2022). The Problem of Reference Rot in Spatial Metadata Catalogues. *ISPRS International Journal of Geo-Information*, 11(1), 27. https://doi.org/10.3390/ijgi11010027

Massicotte, M., & Botter, K. (2017). Reference Rot in the Repository: A Case Study of Electronic Theses and Dissertations (ETDs) in an Academic Library. *Information Technology and Libraries*, 36(1), 11–28. https://doi.org/10.6017/ital.v36i1.9598

Moravcsik, M. J., & Murugesan, P. (1975). Some Results on the Function and Quality of Citations. *Social Studies of Science*, 5(1), 86–92. https://doi.org/10.1177/030631277500500106

Mozilla Foundation. (2025, September 6). *302 Found—HTTP.* https://web.archive.org/web/20250906084515/https://developer.mozilla.org/en-US/docs/Web/HTTP/Reference/Status/302

Nazim, M., Bhardwaj, R. K. (2023). Open access initiatives in European countries: analysis of trends and policies. *Digital Library Perspectives*, 39(3), 371-392. https://doi.org/10.1108/DLP-06-2022-0051

Nelson, M. L., & Allen, B. D. (2002). Object Persistence and Availability in Digital Libraries. *D-Lib Magazine*, 8(1). https://doi.org/10.1045/january2002-nelson

Nicolaisen, J. (2007). Citation analysis. *Annual Review of Information Science and Technology*, 41(1), 609-641. https://doi.org/10.1002/aris.2007.1440410120

Nitsos, I., Malliari, A., & Chamouroudi, R. (2022). Use of reference management software among postgraduate students in Greece. *Journal of Librarianship and Information Science*, 54(1), 95–107. https://doi.org/10.1177/0961000621996413

Phelps, T. A., & Wilensky, R. (2000). Robust Hyperlinks and Locations. *D-Lib Magazine*, 6(7/8). https://doi.org/10.1045/july2000-wilensky

Plantin, J-C., & Thomer, A. (2025). Platforms, programmability, and precarity: The platformization of research repositories in academic libraries. New Media & Society, 27(1), 338-358. https://doi.org/10.1177/14614448231176758

Recker, J., Kleemola, M., & L'Hours, H. (2024). Closing Gaps: A Model of Cumulative Curation and Preservation Levels for Trustworthy Digital Repositories. *International Journal of Digital Curation*, 18(1), 16. https://doi.org/10.2218/ijdc.v18i1.926

Rieger, O. Y. (2012). Sustainability: Scholarly repository as an enterprise. *Bulletin of the American Society for Information Science and Technology*, 39(1), 27–31. https://doi.org/10.1002/bult.2012.1720390110

Romero, R. C. (2025). Transparency in Open Science: An Actionable Principle? *Open Information Science*, 9(1). https://doi.org/10.1515/opis-2025-0016

Rothfritz, L., Matthias, L., Pampel, H., & Wrzesinski, M. (2025). Current challenges and future directions for institutional repositories: A systematic literature review. *Journal of the Association for Information Science and Technology*, n/a(n/a). https://doi.org/10.1002/asi.70016

Rubin, G., Bardi, A., Del Gratta, R., & Del Grosso, A. M. (2026). *Enhancing interoperability of SPARQL endpoints: RESTful and OAI-PMH API for the DH-ATLAS project.* Consiglio Nazionale delle Ricerche, CNR. https://iris.cnr.it/handle/20.500.14243/564781

Russell, E., & Kane, J. (2008). The Missing Link: Assessing the Reliability of Internet Citations in History Journals. *Technology and Culture*, 49(2), 420–429. https://doi.org/10.1353/tech.0.0028

Sauermann, L., & Cyganiak, R. (2008). *Cool URIs for the Semantic Web.* W3C. https://www.w3.org/TR/cooluris/

Schöpfel, J., & Azeroual, O. (2021). Current research information systems and institutional repositories: From data ingestion to convergence and merger. In D. Baker & L. Ellis (Eds), *Future Directions in Digital Information* (pp. 19–37). Chandos Publishing. https://doi.org/10.1016/B978-0-12-822144-0.00002-1

Schöpfel, J., Azeroual, O., Chaudiron, S., Jacquemin, B., Kergosien, E., Prost, H., & Thiault, F. (2024). From Open Repositories to CRIS - A Case Study. *Procedia Computer Science*, 249, 173-178. https://doi.org/10.1016/j.procs.2024.11.061

SCONUL. (2025). *The future of the systems librarian*. SCONUL; https://web.archive.org/web/20250429154425/https://www.sconul.ac.uk/SCONULDownloadController/Download?iType=1&iID=2420&cGUID=29251c00-74a5-4743-bc7a-51827748c4d3&cTempLocation=#expand. https://www.sconul.ac.uk/SCONULDownloadController/Download?iType=1&iID=2420&cGUID=29251c00-74a5-4743-bc7a-51827748c4d3

Shearer, K., Nakano Koga, S. M., Rodrigues, E., Manola, N., Pronk, M., & Proudman, V. (2023). *Current State and Future Directions for Open Repositories in Europe*. COAR, LIBER, OpenAIRE, SPARC Europe, University of Minho. https://doi.org/10.5281/zenodo.10255559

Shearer, K., Rodrigues, E., Manola, N., Malaguarnera, G., Pronk, M., & Proudman, V. (2025). *REPOSITORIES: Key Infrastructure For Maintaining European Research Excellence*. [Zenodo]. https://doi.org/10.5281/ZENODO.15119848

Sherrick, A. K., & Pino Navarro, D. A. (2024, June 28). Creating a better balance: The need for tools and practices to combat AI harvests and resource flooding in repository environments. *Open Repositories 2024 (OR2024)*. https://doi.org/10.5281/zenodo.12579304

Shreeves, S. L., & Cragin, M. H. (2008). Institutional Repositories: Current State and Future. *Library Trends*, *57*(2), 89–97. https://doi.org/10.1353/lib.0.0037

Silva, J. A. T. da. (2021). Reference rot degrades information preservation and induces the loss of intellectual integrity. *Epistēmēs Metron Logos*, 6, 1–6. https://doi.org/10.12681/eml.25792

Speare, M. (2018). Graduate Student Use and Non-use of Reference and PDF Management Software: An Exploratory Study. *The Journal of Academic Librarianship*, 44(6), 762–774. https://doi.org/10.1016/j.acalib.2018.09.019

Spinellis, D. (2003). The decay and failures of web references. *Communications of the ACM*, 46(1), 71–77. https://doi.org/10.1145/602421.602422

Strecker, D., Pampel, H., Schabinger, R., & Weisweiler, N. L. (2023). Disappearing repositories: Taking an infrastructure perspective on the long-term availability of research data. *Quantitative Science Studies*, 4(4), 839–856. https://doi.org/10.1162/qss_a_00277

Suber, P. (2012). *Open Access*. The MIT Press. https://library.oapen.org/handle/20.500.12657/26065

Tait, E., Martzoukou, K., & Reid, P. (2016). Libraries for the future: The role of IT utilities in the transformation of academic libraries. *Palgrave Communications*, 2(1), 16070. https://doi.org/10.1057/palcomms.2016.70

Thalhammer, A. (2024). *Cool URIs for FAIR Knowledge Graphs* (No. arXiv:2407.09237). arXiv. https://doi.org/10.48550/arXiv.2407.09237

Thompson, J., & Hoover, S. (2023). Discoverability Within the Library. In L. Woolcott & A. Shiri, *Discoverability in Digital Repositories* (1st edn, pp. 117–136). Routledge. https://doi.org/10.4324/9781003216438-9

Van de Sompel, H., Klein, M., & Jones, S. M. (2016). Persistent URIs Must Be Used To Be Persistent. *Proceedings of the 25th International Conference Companion on World Wide Web*, 119–120. https://doi.org/10.1145/2872518.2889352

Van de Sompel, H., Nelson, M. L., Sanderson, R., Balakireva, L. L., Ainsworth, S., & Shankar, H. (2009). *Memento: Time Travel for the Web (Version 2)*. arXiv. https://doi.org/10.48550/ARXIV.0911.1112

Van de Sompel, H., Nelson, M., & Sanderson, R. (2013). *HTTP Framework for Time-Based Access to Resource States – Memento (Request for Comments No. RFC 7089)*. Internet Engineering Task Force. https://doi.org/10.17487/RFC7089

van Lieshout, N., Ramezani, S., van Horik, R., Horton, L., Turner, D., Davidson, J., Marjamaa-Mankinen, L., Lager, L., & Nordling, J. (2025). *MS3.8 Technical EOSC PID implementation guide & program (Milestone Report No. 3.8)*. Zenodo. https://doi.org/10.5281/zenodo.14779609

Walk, P. (2023, June 13). Analysis of OpenDOAR Data. *The 18th International Conference on Open Repositories*, Stellenbosch University. https://doi.org/10.5281/ZENODO.8091490

Williams, L., & Woods, L. (2024). Reference management practices of students, researchers, and academic staff. *The Journal of Academic Librarianship*, 50(3), 102879. https://doi.org/10.1016/j.acalib.2024.102879

Yakel, E., Faniel, I. M., Kriesberg, A., & Yoon, A. (2013). Trust in Digital Repositories. *International Journal of Digital Curation*, 8(1), 143–156. https://doi.org/10.2218/ijdc.v8i1.251

Yakel, E., Faniel, I. M., & Robert Jr, L. P. (2024). An empirical examination of data reuser trust in a digital repository. *Journal of the Association for Information Science and Technology*, 75(8), 898–915. https://doi.org/10.1002/asi.24933

Zhou, K., Grover, C., Klein, M., & Tobin, R. (2015). No More 404s: Predicting Referenced Link Rot in Scholarly Articles for Pro-Active Archiving. *Proceedings of the 15th ACM/IEEE-CS Joint Conference on Digital Libraries*, 233–236. https://doi.org/10.1145/2756406.2756940